\documentclass
[pre,twocolumn,showpacs,showkeys,nobibnotes,floatfix]
{revtex4-1}

\usepackage{amsmath,amssymb,array}

\ifx\pdfoutput\undefined
\usepackage{graphicx}
\else \usepackage[pdftex]{graphicx}
\fi

\newcommand\Rey{\ensuremath{\mathrm{Re}}}

\newcommand\Tau{\ensuremath{\tau}}
\renewcommand{\u}{\ensuremath{\mathbf{u}}}
\newcommand{\vv}{\ensuremath{\mathbf{v}}}
\newcommand{\U}{\ensuremath{\mathbf{U}}}
\newcommand{\up}{\ensuremath{\mathbf{u'}}}
\newcommand{\uh}{\ensuremath{\mathbf{\hat u}}}
\newcommand{\x}{\ensuremath{\mathbf{x}}}

\newcommand{\ux}{\ensuremath{u}}
\newcommand{\uy}{\ensuremath{v}}
\newcommand{\uz}{\ensuremath{w}}

\newcommand{\ex}{\ensuremath{\bm{e_x}}}

\newcommand{\A}[1]   	  {\ensuremath{\mathcal{A}(#1)}}

\newcommand{\Astar}[1] 	  {\ensuremath{\mathcal{A}^{\ast}(#1)}}
\newcommand{\AstarA}[1]	  {\ensuremath{\mathcal{A}^{\ast}(#1)\mathcal{A}(#1)}}

\newcommand\umode{\tilde{\bm{u}}}
\newcommand\uadj{\bm{u}^*}

\DeclareMathOperator*{\argmax}{arg\,max}

\newcommand\etal{{\it et al}.}
\newcommand{\ie}{i.e.\ }
\newcommand{\eg}{e.g.\ }

\newcommand\BvK{B\'enard-von K\'arm\'an}

\newcommand\Mopt {\Omega_{\text{L}}}
\newcommand\Msmall {\Omega_{\text{S}}}
\newcommand\Bsmall {\U_{\text{S}}}
\newcommand\Bopt {\U_{\text{L}}}

\newcommand\Psmall {\up_{\text{S}}}
\newcommand\Popt {\up_{\text{L}}}

\begin{document}

\title{Computational study of subcritical response in flow past a circular 
cylinder}
\author{C. D. Cantwell}
\email[Email: ]{c.cantwell@imperial.ac.uk}
\author{D. Barkley}
\email[Email: ]{D.Barkley@warwick.ac.uk} \affiliation{Mathematics Institute and
Centre for Scientific Computing, University of Warwick, 
Coventry CV4 7AL, United Kingdom}

\date{\today}

\begin{abstract}

Flow past a circular cylinder is investigated in the subcritical regime, below
the onset of B\'enard-von K\'arm\'an vortex shedding at $\Rey_c \simeq 47$.
The transient response of infinitesimal perturbations is computed.  The domain
requirements for obtaining converged results is discussed at length.  It is
shown that energy amplification occurs as low as $\Rey=2.2$.  Throughout much
of the subcritical regime the maximum energy amplification increases
approximately exponentially in the square of $\Rey$ reaching $6800$ at
$\Rey_c$.  The spatiotemporal structure of the optimal transient dynamics is
shown to be transitory B\'enard-von K\'arm\'an vortex streets.  At
$\Rey\simeq42$ the long-time structure switches from exponentially increasing
downstream to exponentially decaying downstream.  Three-dimensional
computations show that two-dimensional structures dominate the energy growth
except at short times.

\end{abstract}




\maketitle

\section{Introduction}

Incompressible fluid flow past a circular cylinder has been extensively
studied, both for its relevance to numerous engineering applications and as a
prototype bluff-body flow exhibiting vortex shedding. See for
example~\cite{Zdravkovich:1997, Zdravkovich:2003} and references therein.  It
is one of the most-used test bed for exploring stability concepts in
open flows, e.g.~\cite{ProvansalMathisBoyer:1987, Jackson:1987, Zebib:1987, 
  YangZebib:1989, Dusek:1994, NoackEckelmann:1994, Pier:2002, Noack:03,
  Chomaz:2005, Barkley:2006, GiannettiLuchini:2007}.
As a result, a great deal is known about this flow in general and
in particular concerning the primary instability.
It is well-established that this instability occurs at a critical Reynolds
number of about 47~\cite{ProvansalMathisBoyer:1987, Jackson:1987, Zebib:1987}.
Below this value the steady wake flow is linearly stable, while above it the
steady flow is unstable and periodic oscillations arise leading to the famous
\BvK\ vortex street~\cite{Benard:1908, VonKarman:1911}. Our concern
here is what happens in the stable, subcritical regime prior to, and leading up
to, the onset of oscillations.

Stable flows may exhibit transient growth~\cite{SchmidHenningson:2001,
  TrefethenEmbree:2005}. This means that infinitesimal perturbations to the
flow may grow in energy for some time before subsequently decaying to zero.
While initially popular in parallel shear flows as possibly playing a role in
the transition to turbulence, e.g.~\cite{Gustavsson:1991, Butler:1992,
  Henningson:1993}, transient growth has become increasingly of interest in
spatially developing flows, e.g.~\cite{CossuChomaz:1997,Chomaz:2005,
  AAkervikHoepffnerEhrensteinHenningson:2007, BlackburnBarkleySherwin:2008,
  MarquetSippChomazJacquin:2008, Alizard:2009}. For instance, separated flows
arising due to abrupt changes in geometry are known to promote extremely large
transient growth in perturbations~\cite{BlackburnBarkleySherwin:2008,
  BlackburnSherwinBarkley:2008, CantwellBarkleyBlackburn:2010}.  The origin of
this growth can be traced to the non-normality of the linear stability
operator associated with many shear
flows~\cite{Chomaz:2005,SchmidHenningson:2001}.  This means, in particular,
that in spatially developing flows the eigenmodes of the stability operator
tend to be located downstream while the eigenmodes of the adjoint operator
tend to be located upstream~\cite{TrefethenEmbree:2005, Chomaz:2005, 
GiannettiLuchini:2007}.

For the cylinder wake, Giannetti and Luchini~\cite{GiannettiLuchini:2007} first
examined in detail the adjoint eigenmodes in the vicinity of the primary
instability and used these, together with direct eigenmodes, to understand the
sensitivity of the flow. Their results are further discussed in detail by
Chomaz in the context of non-normality~\cite{Chomaz:2005}.  Since this important
work, there have been further computations of direct and adjoint modes and
transient growth for the cylinder wake. For example Marquet
\etal~\cite{MarquetSippLaurent:2008} have computed direct and adjoint
eigenmodes of the supercritical flow, and Abdessemed
\etal~\cite{AbdessemedSharmaSherwinTheofilis:2009} have studied the transient
growth, focusing on supercritical Reynolds numbers, although also reporting
some subcritical values.

There have been a number of experimental studies of the cylinder wake in the
stable and marginally unstable regime~\cite{ProvansalMathisBoyer:1987,
  LeGalCroquette:2000,Maraisetal:2010}.  The most relevant are the studies by
Le Gal and Croquette~\cite{LeGalCroquette:2000} and the recent work by Marais
\etal~\cite{Maraisetal:2010} on the impulse response at subcritical Reynolds
numbers. By inducing an impulse through a small displacement or rotation of
the cylinder, wavepackets are generated that grow and are subsequently
advected downstream. While the measurements by Le Gal and Croquette provide
informative qualitative properties of the transient dynamics, these
measurements were made using streaklines and so provide limited quantitative
detail. The more recent work by Marais \etal\ uses particle image velocimetry
to obtain quantitative measurements of the transient response in the
subcritical regime.

The purpose of the current study is two-fold. Primarily we establish an
accurate characterization of the optimal transient energy growth throughout
the subcritical regime for the cylinder wake. We determine the threshold
Reynolds number where energy growth first occurs, determine the Reynolds
number dependence of the optimal growth, and its value at criticality. We
show that the transient dynamics associated with optimal energy growth is in
the form of wave packets similar to those observed in experiments on
subcritical wakes. The secondary purpose of the paper is to highlight and
establish the computational requirements for such computations. As we shall
show, the requirements for accurate computations of transient growth are more
severe than those of linear stability. While these findings are specific to
the cylinder wake, they should guide computations of similar flows.
\section{Formulation}
\label{sec:formulation}

\begin{figure}
\centering
\includegraphics{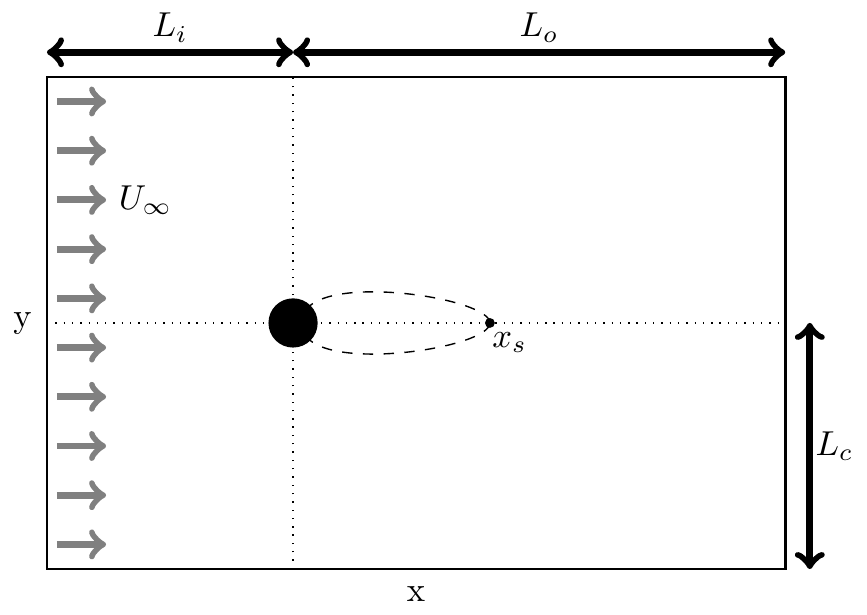}
\caption[Diagram of the cylinder geometry.]
{Diagram of the cylinder geometry (not to scale), showing
the inflow, outflow and cross-stream dimensions referenced later. Also marked
are the separation streamlines and the downstream stagnation point, $x_s$.}
\label{f:cyltg:geometry}
\end{figure}

The flow geometry is illustrated in Fig.~\ref{f:cyltg:geometry}.  A circular
cylinder of diameter $D$ is placed in a free-stream flow
$U_{\infty}\ex$. Streamwise $x$ and cross-stream $y$ coordinates are
centered on the circular cross section.  The cylinder axis, infinite in length
and normal to the free-stream velocity, aligns with the $z$-coordinate.

In principle this open flow would have infinite extent in all directions. In
practice, however, our numerical calculations necessarily employ a
computational domain $\Omega$ with finite inflow $L_i$, outflow $L_o$, and
cross-stream $L_c$ lengths, as illustrated. The $z$-direction is homogeneous,
and for the issues addressed in this paper, this direction can be treated
without needing to restrict to a bounded domain.  The demands on the domain
dimensions is an important aspect of our work discussed in detail in
Sec.~\ref{sec:cyltg:mesh}.

The fluid is governed by the incompressible Navier-Stokes equations
\begin{subequations}
	\label{e:cyltg:nse}
	\begin{align}
		\partial_t \u + (\u \cdot \nabla)\u &= -\nabla p 
                + \Rey^{-1} \nabla^2 \u, \\
		\nabla \cdot \u &= 0,
	\end{align}
\end{subequations}
where $\u = \u(\x) = (\ux(x,y,z),\uy(x,y,z),\uz(x,y,z))$ is the fluid velocity
and $p(x,y,z)$ is the static pressure. Without loss of generality we set the
density to unity. The equations are non-dimensionalized by the free-stream
speed $U_{\infty}$ and the cylinder diameter $D$. The Reynolds number $\Rey$
is therefore given as
\begin{align*}
	\Rey = \frac{U_{\infty}D}{\nu},
\end{align*}
where $\nu$ is the kinematic viscosity of the fluid. 

No-slip boundary conditions are imposed on the cylinder surface. The boundary
conditions around the outer boundaries of the domain are such as to give a
good numerical approximation of the unbounded flow. Specifically, the boundary
conditions are:
\begin{subequations}
	\label{eq:nse-bc}
	\begin{align}
	\label{eq:nse-bc-in}
    	\bm{u}(\partial\Omega_i,t) &= U_{\infty} \ex,\\
    	\bm{u}(\partial\Omega_c,t) &= U_{\infty} \ex,\\
    	\bm{u}(\partial\Omega_w,t) &= \bm{0},\\
	\label{eq:nse-bc-out}
    	\partial_x \bm{u}(\partial\Omega_o,t) &= 0, 
        \quad p(\partial\Omega_o,t) = 0,
    \end{align}
\end{subequations}
where $\partial\Omega_i$ is the inlet boundary at $x=-L_i$, 
$\partial\Omega_c$ is the cross-stream boundary at $y=\pm L_c$,
$\partial\Omega_w$ is the boundary of the cylinder,
and $\partial\Omega_o$ is the outlet boundary at $x=L_o$.

The remaining material in this section is included for completeness and to
clearly define notation. Since the details are contained in numerous prior
publications,
especially~\cite{TuckermanBarkley:2000,BarkleyBlackburnSherwin:2008}, the
treatment here is minimal.

Equation~\eqref{e:cyltg:nse} is solved using Direct Numerical Simulation
(DNS) employing a split-step pressure-correction scheme described elsewhere
\cite{OrszagIsraeliDeville:1986,KarniadakisIsraeliOrszag:1991}.  This is
implemented in a spectral-element code~\cite{BlackburnSherwin:2004} utilizing an
elemental decomposition of the domain in the two-dimensional (2D) plane normal
to the cylinder axis.

\begin{figure}
\centering
\includegraphics{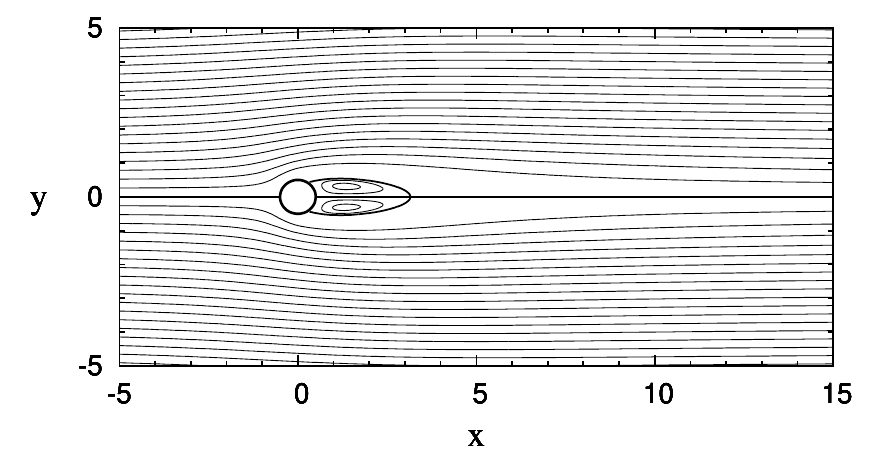}
\caption
{Representative base flow. Streamlines of the flow at \Rey=40. The
  computational domain is much larger than the region shown. The main flow
  field uses a contour spacing of 0.25, while a smaller spacing of 0.01 is
  used to highlight the structure of the recirculation bubble. }
\label{f:cyltg:stream}
\end{figure}

The base flows $\U$ considered in this paper are steady, two-dimensional
solutions to Eq.~\eqref{e:cyltg:nse}. Hence $\U = (U(x,y),V(x,y))$.  These
Reynolds number dependent flows are symmetric about the streamwise centerline
as depicted in Fig.~\ref{f:cyltg:geometry}.  
Figure~\ref{f:cyltg:stream} shows a typical base flow.  Those at other $\Rey$
are qualitatively similar, differing primarily in the length of the
recirculation region behind the cylinder.  For $\Rey \lesssim 6.2$, there is
no recirculation region.
Steady base flows in both
subcritical and supercritical regimes are rapidly obtained through DNS by
imposing this midplane symmetry.  Once computed, base flows are stored for use
in subsequent linear calculations.

Our interest is in the dynamics of infinitesimal perturbations $\up$ to the
steady base flow. These perturbations evolve according to the linearized
Navier-Stokes equations
\begin{subequations}
	\label{e:cyltg:lnse}
	\begin{align}
		\partial_t \up + (\up \cdot \nabla) \U + (\U \cdot \nabla) \up 
	        &= -\nabla p' + \Rey^{-1} \nabla^2 \up, \\
		\nabla \cdot \up &= 0,
	\end{align}
\end{subequations}
where $p'$ is the perturbation pressure.  Numerically
Eq.~\eqref{e:cyltg:lnse} is solved using the same techniques as the
nonlinear Navier-Stokes equations. For the most part we will focus on 2D
perturbation fields $\up = (u', v')$ on 2D grids. However, we will also
consider briefly three-dimensional perturbations. Since the base flow is 2D,
three-dimensional (3D) perturbations can be decomposed into non-interacting
modes of the form
\begin{equation}
\up(x,y,z) = \uh(x,y) e^{i\beta z} + c.c.,
\label{eq:3Dpert}
\end{equation}
where $\beta$ is the spanwise wavenumber of the perturbation. Only $\uh(x,y)$, 
a three-component field on a two-dimensional grid, needs to be computed.

Our primary focus is on the transient dynamics of perturbations at subcritical
Reynolds numbers.  We focus on the energy of perturbation fields and seek
initial conditions $\up(0)$ which generate the largest possible growth in
energy under evolution by Eq.~\eqref{e:cyltg:lnse}. The formalism is as
follows.  Let \A{\tau} denote the linear evolution operator over a time \Tau\,
defined by Eq.~\eqref{e:cyltg:lnse}, so that
\begin{align*}
	\up(\tau) = \A{\tau}\up(0).
\end{align*}

Since the governing equations are linear it is sufficient to consider initial
perturbation fields with $||\up(0)||^2 = \langle\up(0),\up(0)\rangle = 1$,
where $\langle\cdot,\cdot\rangle$ denotes the $L_2$ inner-product.
Then the energy growth in the perturbation field over a time $\tau$ is given
by
\begin{align*}
\frac{E(\tau)}{E(0)} = \langle\up(\tau),\up(\tau)\rangle.
\end{align*}
In terms of the operator \A{\Tau}, and its adjoint \Astar{\Tau} in the $L_2$
inner-product, we have
\begin{align*}
\frac{E(\tau)}{E(0)} = & \langle\A{\tau}\up(0),\A{\tau}\up(0)\rangle \\
                     = & \langle\up(0),\AstarA{\tau}\up(0)\rangle.
\end{align*}

Letting $\lambda_j$ and $\vv_j$ denote eigenvalues and normalized
eigenfunctions of the operator \AstarA{\tau}, we have
\begin{equation}
\label{eq:A*A_ev}
\AstarA{\tau} \vv_j = \lambda_j \vv_j, \qquad ||\vv_j|| = 1.
\end{equation}
The eigenvalues are non-negative and we assume ordering $\lambda_1 \ge
\lambda_2\ge \cdots$. 

The maximum possible energy growth, denoted $G(\tau)$, over a specified time
horizon \Tau, is then given by the dominant eigenvalue of $\AstarA{\tau}$, \ie
\begin{align*}
    G(\tau) = \max_j \lambda_j = \lambda_1,
\end{align*}
The initial perturbation leading to this growth is the corresponding
eigenfunction $\vv_1$.  While the dominant eigenvalue of $\AstarA{\tau}$ is
generally of most importance, the first few sub-dominant eigenvalues may also
be of interest. In particular $\vv_2$ will also be considered in this study. 

The maximum energy growth over all time horizons is denoted by
\begin{align}
	G^{\max} = G(\tau^{\max}),  
        \label{eq:gmax}
\end{align}
where
\begin{align}
	\tau^{\max} = \argmax_{\tau} G(\tau).
        \label{eq:taumax}
\end{align}

While our primary focus is transient growth, we report some eigenvalue
results. Equation~\eqref{e:cyltg:lnse} can formally be written 
\begin{align*}
	\partial_t{\up} = \mathcal{L}\up,
\end{align*}
Looking for normal-mode solutions to these equations gives the eigenvalue
problem 
\begin{align}
  \mathcal{L} \umode_j = \sigma_j \umode_j, \qquad ||\umode_j|| = 1,
  \label{eq:evprob}
\end{align}
where $\umode_j$ are normalized eigenmodes and $\sigma_j$ eigenvalues of
$\mathcal{L}$. We assume ordering such that ${\rm Re}(\sigma_1) \ge {\rm
  Re}(\sigma_2) \ge \cdots$.  Stability of the base flow is determined from
the right-most eigenvalues of $\mathcal{L}$ in the complex plane.

Associated to Eq.~\eqref{eq:evprob} is the adjoint eigenvalue
problem
\begin{align}
  \mathcal{L}^* \uadj_j = \sigma_j^* \uadj_j, 
  \label{eq:adjprob}
\end{align}
where $\uadj_j$ are the adjoint modes, (eigenmodes of the adjoint operator
$\mathcal{L}^*$), and $\sigma_j^*$ is the complex conjugate of $\sigma_j$.
The norm of the adjoint eigenmodes is chosen so that $\langle \uadj_j,
\umode_j\rangle = 1$ for all $j$. Then the eigenmode and adjoint modes satisfy
biorthonormality:
\begin{align*}
\langle \uadj_i, \umode_j\rangle = \delta_{ij}.
\end{align*}

In practice the eigenvalues for the transient growth or stability problems are
computed through a modified Arnoldi algorithm using a time-stepper
approach~\cite{TuckermanBarkley:2000,BarkleyBlackburnSherwin:2008}.

\section{Influence of Domain Size}
\label{sec:cyltg:mesh}

As noted in the introduction, the size of the computational domain can be an
important factor in studies of the transient response in subcritical cylinder
flow. 
While the requirements for accurate base flows and eigenvalue calculations for
the cylinder wake have been discussed in many places
\cite{Fornberg:1980,LecointePiquet:1984,StrykowskiHannemann:1991,
AnagnostopoulosIliadisRichardson:1998}, and are presented in our study in the
Appendix, there is no such discussion for transient growth calculations for
the cylinder wake. Hence some details are worthwhile. We first present
results from the convergence study and then discuss some of the causes and
implications of our findings.

\subsection{Convergence}
\label{sec:cyltg:mesh:tg}

We focus on the role of inflow length, $L_i$, and the cross-stream
half-length, $L_c$, since these are the critical lengths. The requirements on
the outflow length, $L_o$, are set by the largest $\tau$ value under
consideration in the transient growth analysis and could in principle be
arbitrarily large.  Based on the maximum value of $\tau=110$ we consider, and
a free-stream $U_{\infty}=1$, we fix the outflow length in the convergence
study at $L_o=125$.

\begin{figure}
\centering
\includegraphics[scale=0.9]{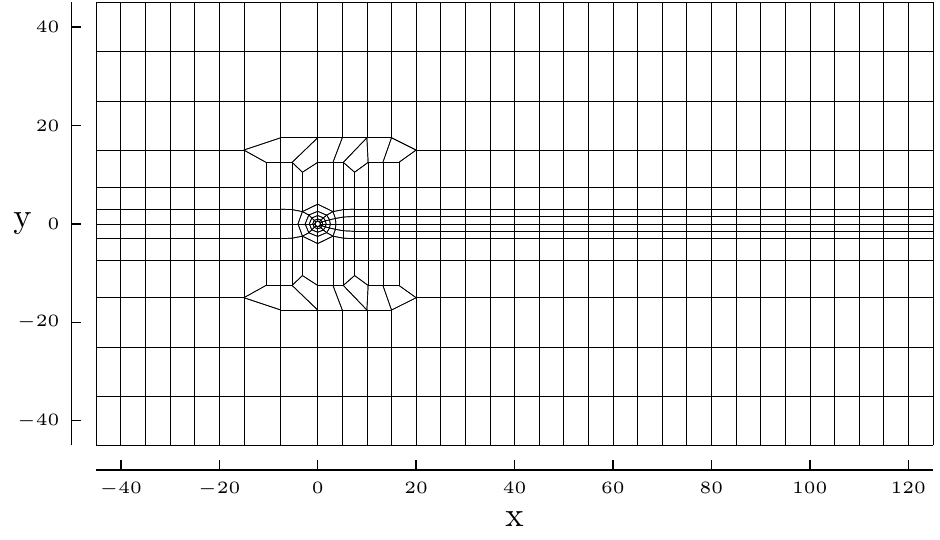}
\caption
{Representative spectral-elemental mesh. Dimensions are $L_i=45$, $L_c=45$ and
$L_o=125$ (refer to Fig.~\ref{f:cyltg:geometry} for definitions).}
\label{f:cyltg:mesh}
\end{figure}

Figure~\ref{f:cyltg:mesh} shows a representative spectral-element domain of
the type used in our computations. (It is in fact the final mesh used for
obtaining results presented in Sec.~\ref{sec:cyltg:tg}.) For the study of
domain size, $L_i$ and $L_c$ are varied by adding or removing elements as
necessary, and the polynomial order of the spectral expansion within each
element is fixed at order 8. The polynomial order used in obtaining the final
transient growth results is $6$, as established in the Appendix.

\begin{figure}
\centering
\includegraphics{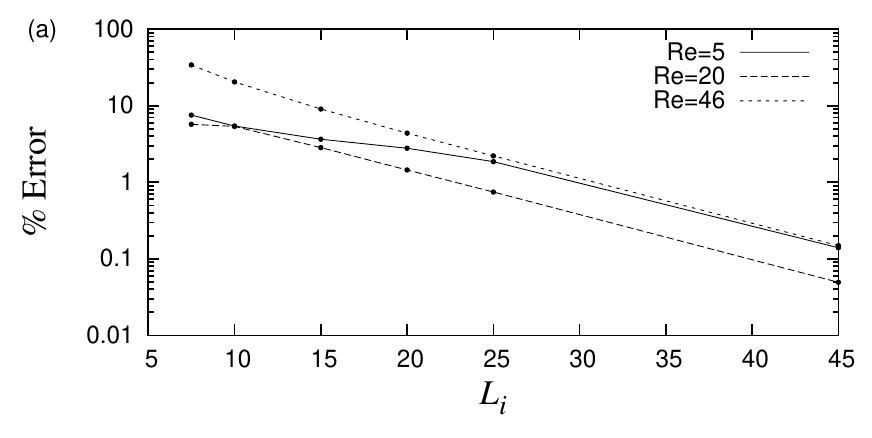}\\
\includegraphics{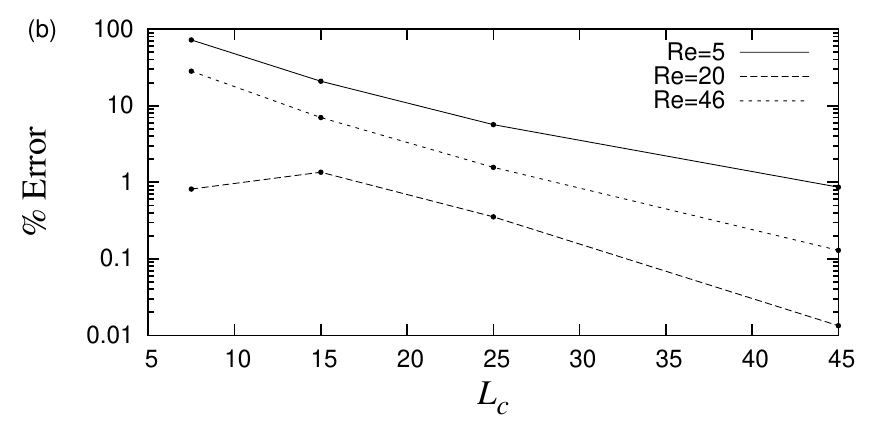}
\caption{Convergence of optimal growth calculation with mesh geometry by (a)
  in-flow length $L_i$ and (b) cross-stream length $L_c$. 
  Points indicate the computed values. Optimal growth is
  for a time horizon of $\tau=20$. Percentage errors are shown relative to the
  calculation using $L_i=65$ and $L_c=65$, respectively.}
\label{f:cyltg:converge-mesh}
\end{figure}

The dependence of these calculations on domain size is assessed through the
calculation of energy growth at a fixed time horizon of $\tau=20$. At this
time horizon non-negligible growth is expected across most of the range of
Reynolds numbers under consideration. 
Figure~\ref{f:cyltg:converge-mesh}
summarizes the errors introduced through domain size restriction.
Three Reynolds numbers, $\Rey=5$, $\Rey=20$ and $\Rey=46$, are considered
to ensure that the mesh is capable of resolving all solutions in the
subcritical range.

The transient growth results are seen to be sensitive to domain size, much
more so than either the base flow or eigenvalue calculations presented
in the appendix.
Domains that provide results accurate to within 1\% for base flows and
eigenvalues, e.g.\ a domain with $L_i=L_c=25$, do not provide such accuracy
for transient growth calculations. The effect of cross-stream restriction at
low \Rey\ is particularly significant.  Even accepting that in many cases one
does not need high accuracy in transient growth values,
Fig.~\ref{f:cyltg:converge-mesh} demonstrates the care that must be taken in
computing transient response in the subcritical regime.

Based on these results, a computational domain with $L_i=45$ and $L_c=45$ is
deemed sufficient to resolve transient growth calculations to within about 1\%
for subcritical Reynolds numbers. Possibly the accuracy is not quite 1\% at
$\Rey<5$, but the growth values are sufficient for our purposes.  A
diagram of the resulting mesh is shown in Fig.~\ref{f:cyltg:mesh}.

\subsection{Discussion}

We begin by recalling that recently,
Abdessemed {\em et al.}~\cite{AbdessemedSharmaSherwinTheofilis:2009}
reported transient growth calculations in the cylinder wake, including some
within the subcritical regime at $\Rey=45$. In comparing those results with
ours, we have found that their growth values are about 32\% larger than those
computed on the mesh in Fig.~\ref{f:cyltg:mesh}, at the same \Rey\ and time
horizon.  We will use this discrepancy to focus the present discussion.

The Abdessemed {\em et al.} calculations were performed using a
spectral-element code similar to that used in this study.  Their computational
domain has bounds $-8\leq x\leq 95$ and $-12.5\leq y\leq 12.5$ and identical
boundary conditions to ours.  One can quickly rule out the possibility
that resolution (polynomial order) or outflow length are significant factors
in the disagreement between the two computations.  Moreover, we have already
seen that the transient growth calculations require large inflow $L_i$ and
cross-stream $L_c$ dimensions so the disagreement is not surprising in
retrospect. However, there are in fact two causes for the discrepancy which we
address: one is the indirect effect caused by differences in the base flows for
the different computations and the other is the direct effect of domain
requirements for the optimal initial condition itself.

We shall refer to our computational domain with dimensions as in
Fig.~\ref{f:cyltg:mesh} as $\Mopt$, (large domain), and that with dimensions
used by Abdessemed {\em et al.} as $\Msmall$, (small domain).  Let $\Bopt$
denote the base flow computed on $\Mopt$ (at $\Rey=45$) and let $\Popt$ denote
the normalized initial condition, $||\Popt||=1$, giving optimal growth at
$\tau=100$.  Similarly let $\Bsmall$ and $\Psmall$ denote the base flow and
normalized optimal initial condition on $\Msmall$.  The resulting energy
growth for the two calculations is given in the first two rows of
Table~\ref{t:cyltg:results:mesh}, where one sees the large discrepancy.
It is worth pointing out that the critical Reynolds numbers obtained on the
two meshes differ by only about 2\%. 

To assess the role of the base flow, one can take the initial condition
$\Psmall$ from the smaller domain and evolve it on the larger domain with the
corresponding base flow $\Bopt$. The resulting growth after 100 time units is
given in the third row of Table~\ref{t:cyltg:results:mesh}.  Necessarily the
growth had to be less than for $\Popt$ because $\Popt$, by definition, gives the
largest possible growth over this time horizon on $\Mopt$. It is perhaps
somewhat surprising, however, that the growth following from the fixed initial
condition $\Psmall$ is approximately factor of 2 less on the large domain than
on the small domain (second and third rows of
Table~\ref{t:cyltg:results:mesh}).  The difference is almost entirely
attributable to the difference in the base flows $\Bopt$ and $\Bsmall$.  This
is confirmed by evolving $\Psmall$ on the small domain but with the base
flow $\Bopt$, truncated onto the smaller domain. The result is given in the
last row of Table~\ref{t:cyltg:results:mesh}. There is little difference
between the evolution of $\Psmall$ on the two domains, if they both have
the same base flow $\Bopt$.  The conclusion is that the energy growth may
depend considerably on the base flow (a factor of 2 in this case), even in
situations where other measures, such as critical Reynolds numbers, would not
reveal such a large dependency.

\begin{figure}
\centering
\includegraphics{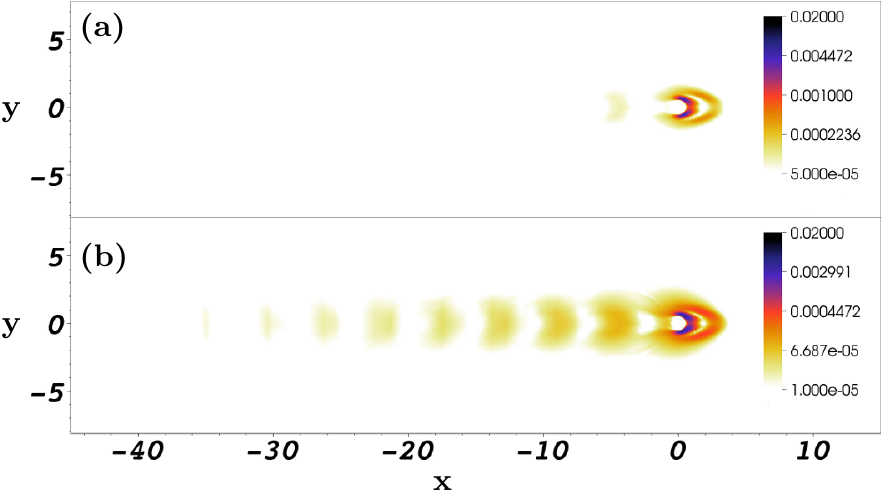}
\caption
{(color online) Energy of the optimal initial condition for $\Rey=45$,
$\Tau=100$ shown on two scales: (a) matching that of the figures shown later 
in the paper and, (b) showing an extended scale which highlights the upstream
tail of the perturbation.}
\label{f:cyltg:opt-ic-inflow}
\end{figure}

There is then the remaining issue of how $\Popt$ and $\Psmall$ differ and why
the energy growth following from $\Psmall$ is 27\% less than from $\Popt$ for
the same base flow (first and third rows in Table~\ref{t:cyltg:results:mesh}).
This has to do with the domain requirements, in particular the inflow length
$L_i$ needed to capture the optimal initial condition. In
Fig.~\ref{f:cyltg:opt-ic-inflow} we show the upstream extent of $\Popt$ on two
scales. The energy of the perturbation upstream of $x=-5$ is of the order
$10^{-5}$ and, while one might consider it to be negligible, this portion of
the initial condition makes a significant contribution to the overall growth
and cannot be neglected in the transient growth computations if quantitative
accuracy is required.

We conclude with a few further remarks on the presence of weak upstream tails
in the optimal initial conditions.  First, despite our caution about the need
to resolve these to obtain quantitatively accurate results, we find the linear
evolution from the optimal initial condition is {\em qualitatively} similar
whether or not the numerical domain fully contains the weak upstream tail of
the initial condition. We observe no important qualitative errors in
discounting it, but quantitatively the errors in the energy growth can be
large. In addition, the length of the upstream tail depends on the time
horizon \Tau. The value \Tau=100 considered in our comparison is rather large.
For smaller time horizons the weak tail may be absent from the optimal initial
condition simply because such a tail could not advect downstream and come into
play over a small time horizon. Specifically, in Sec.~\ref{sec:st} we focus on
optimal initial conditions computed for \Tau=20 and in this case the upstream
tails are absent.

\begin{table}
\begin{ruledtabular}
\begin{tabular}{cccc}
domain	& 	base flow	& 	IC		& $E(100)$ \\
\hline \\
$\Mopt$	   & 	$\Bopt$		& 	$\Popt$		& $2.54\times 10^3$ \\
$\Msmall$  &	$\Bsmall$	& 	$\Psmall$	& $3.36\times 10^3$ \\
$\Mopt$	   &	$\Bopt$		&	$\Psmall$	& $1.85\times 10^3$ \\
$\Msmall$  &	$\Bopt$	        &	$\Psmall$	& $1.77\times 10^3$ \\
\end{tabular}
\end{ruledtabular}
\caption
{Effects of domain size, base flow, and initial condition (IC) on the energy
  growth at $\Rey=45$.  Energy at time $100$ is given for different
  configurations (see text). }
\label{t:cyltg:results:mesh}
\end{table}

\section{Transient Subcritial Response}
\label{sec:cyltg:tg}

\subsection{2D Energy Growth}

\begin{figure}
\centering
\includegraphics{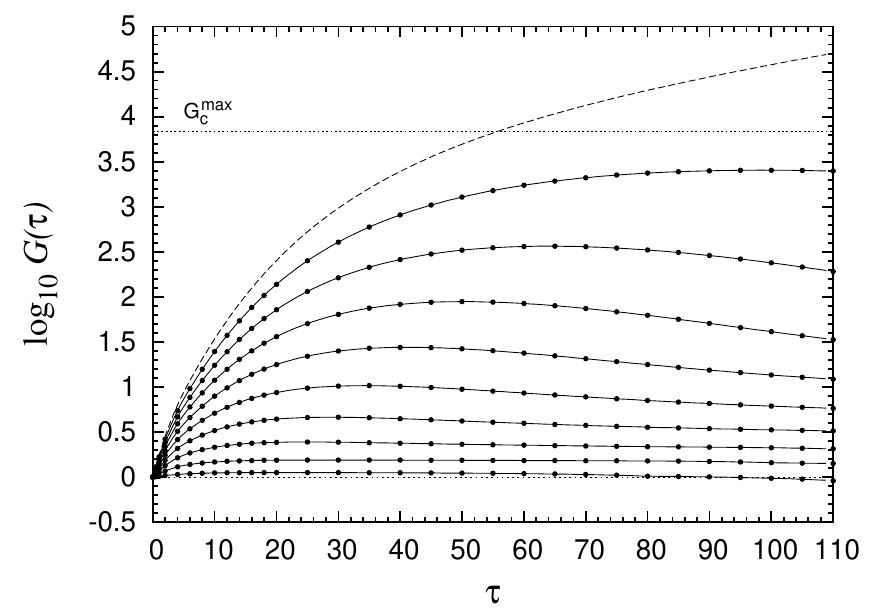}
\caption{Optimal energy growth at Reynolds numbers from $\Rey=5$ to $\Rey=50$
  in increments of $\Rey=5$. Points indicate the computed values. The case
  $\Rey=50$ is above $\Rey_c$ and is shown as dashed.  The horizontal line is
  an estimate of the maximum growth in the subcritical regime (see text).}
\label{f:cyltg:growth}
\end{figure}

\begin{figure}
\centering
\includegraphics{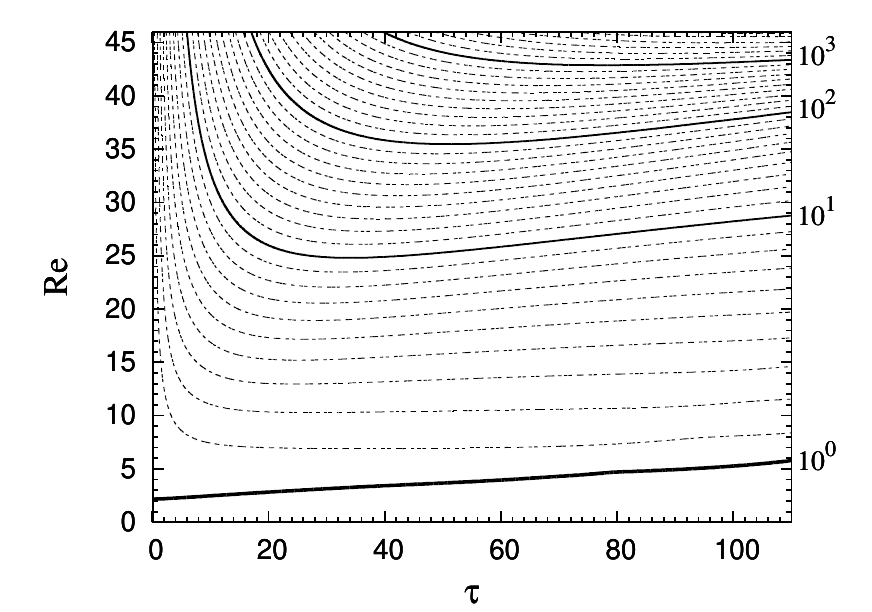}
\caption{Contour plot of optimal energy growth in the subcritical regime, with
  contour levels as indicated. The thick black curve denotes the contour of
  no-growth, $G=1$.}
\label{f:cyltg:growth-contour}
\end{figure}

Figures~\ref{f:cyltg:growth} and \ref{f:cyltg:growth-contour} summarize the
optimal energy growth for 2D perturbations in the subcritical regime.
Figure~\ref{f:cyltg:growth} shows the optimal growth envelopes for particular
values of \Rey. To be clear, these curves show the largest attainable energy
growth over all possible initial conditions at each value of $\tau$.  The
uppermost curve is the growth at $Re=50$, above the onset of linear
instability at $\Rey_c$.  After an initial rapid growth, the energy increase
saturates to an exponential rate, in line with that of the leading eigenvalue.

Figure~\ref{f:cyltg:growth-contour} shows growth contours in the (\Rey,
$\tau$) plane.  The contours highlight the fast energy growth at small time
horizons and the slow decay for long \Tau.  The thick curve denotes the
no-growth contour: $G=1$.  The interception of this curve with the \Rey-axis
indicates a critical Reynolds number for energy growth~\cite{Joseph:1976},
$\Rey_E$, below which all perturbations decay monotonically in time and above
which there is transient energy growth for at least some perturbations.  We
estimate $\Rey_E=2.2$. Thus, a small amount of transient energy growth is
possible before the formation of the recirculation region behind the cylinder
at $\Rey = 6.2$.

\begin{figure}
\centering
\includegraphics{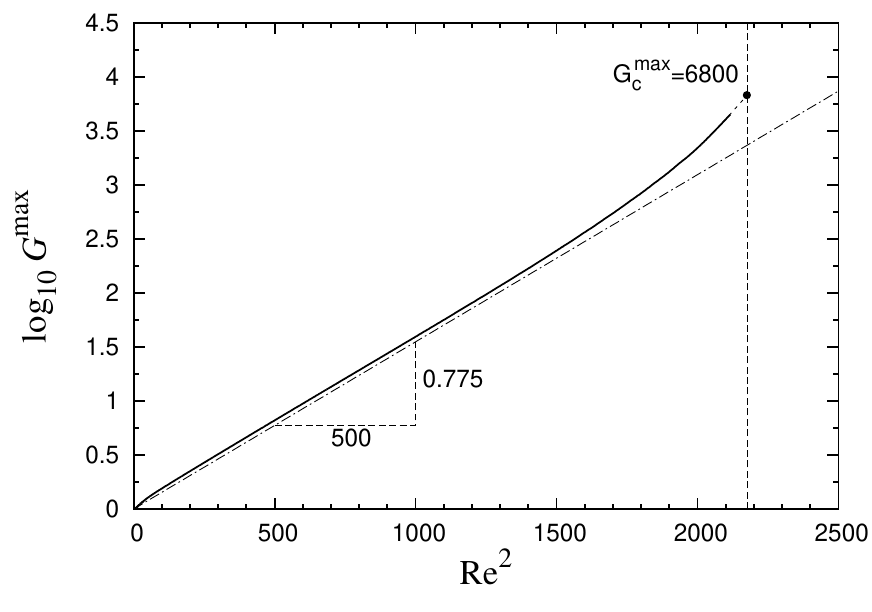}
\caption{Maximum growth $G^{max}$ as a function of the
  square of Reynolds number. The vertical dotted line corresponds to $\Rey_c$.
  $G^{max}_c$, the maximum growth at $\Rey_c$, is indicated with a point and is
  computed differently from cases $\Rey \le 46$ (see text).  The dashed-dotted
  line highlights the relationship given by Eq.~\eqref{eq:gmaxform}.  }
\label{f:cyltg:growth-max}
\end{figure}

The single most important measure of the transient energy growth at any
\Rey\ is the maximum $G^{\max}$ over all $\tau$. [Recall Eqs.~\eqref{eq:gmax}
and \eqref{eq:taumax}].  This is shown in Fig.~\ref{f:cyltg:growth-max} where
$\log_{10} G^{\max}$ is plotted as a function of $\Rey^2$.  
Throughout most of the
subcritical regime, the maximum growth increases exponentially with
$\Rey^2$. More specifically, we find
\begin{equation}
G^{\max} \simeq \exp(1.55\times 10^{-3} \Rey^2).
\label{eq:gmaxform}
\end{equation}
Only for $\Rey \gtrsim 40$ does the growth deviate significantly from this
form. There is an upturn in the maximum growth in approaching $\Rey_c$.  Above
$\Rey_c$, $G^{\max} = \infty$ since the flow is linearly unstable and
$G(\tau)$ diverges as $\tau \to \infty$.

Data up to $\Rey=46$ have been obtained via the transient growth calculations
described in Sec.~\ref{sec:formulation}. The maximum growth at $\Rey_c$ is
obtained differently.  At criticality, the optimal initial condition is the
adjoint eigenmode corresponding to the critical eigenvalue. Under linear
evolution, this initial condition evolves to the direct eigenmode. Hence the
optimal growth at criticality is obtained from the eigenmode and its adjoint.
Taking $||\tilde u|| =1$ and $\langle u^*, \tilde u \rangle =1$, \ie
biorthonormalized modes, then the maximum growth is given by $||u^*||$.  Based
on these calculations our estimate of the maximum growth at $\Rey_c$, and
hence the maximum growth within the subcritical regime, is $G^{\max}_c \equiv
G^{\max}(\Rey_c) \approx 6800$.

\subsection{Spatiotemporal Evolution}
\label{sec:st}

We now turn to one of our primary focuses, the transient evolution of
infinitesimal perturbations. For the most part we shall be interested
in the qualitative character of this evolution and how it depends on $\Rey$.

\begin{figure}
\centering
\includegraphics{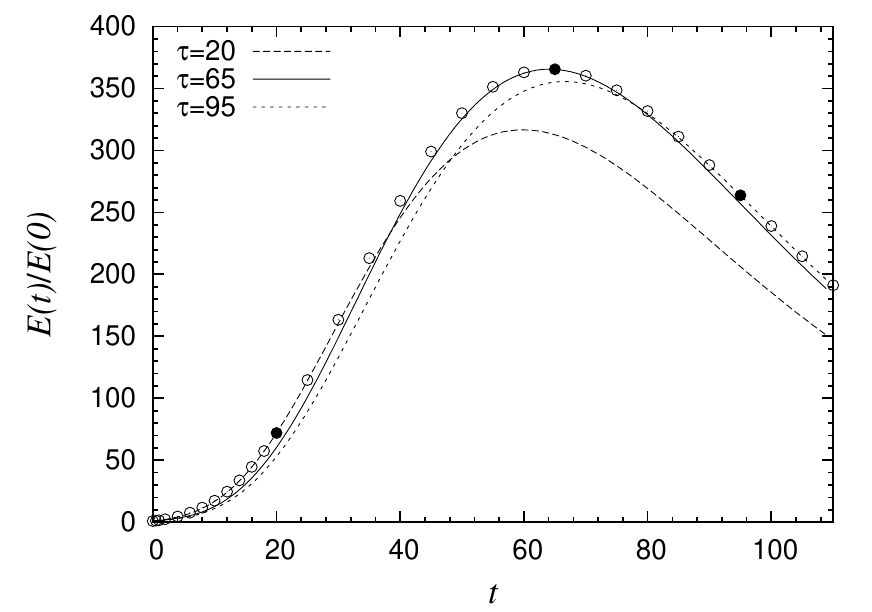}
\caption{Energy of evolving perturbations computed at \Rey=40 for three time
  horizons: \Tau=20, \Tau=65, and \Tau=95. The three curves touch the optimal
  growth envelope (circles) at their respective $\tau$ values. Qualitatively
  similar evolution is seen over a large range of optimal perturbations.}
\label{f:cyltg:growth-time}
\end{figure}

We start the perturbation field $\up$ from initial conditions calculated for
optimal linear energy growth and evolve the field via the linearized
Navier-Stokes equations, Eq.~\eqref{e:cyltg:lnse}.  Recall from
Sec.~\ref{sec:formulation} that the initial condition giving optimal growth
over time horizon \Tau\ is the dominant eigenfunction $\vv_1$ of
$\AstarA{\tau}$.  Hence, the initial condition and subsequent evolution
$\up(t)$, depend on time horizon $\tau$ in defining $\AstarA{\tau}$.  However,
the transient dynamics based on a substantial range of $\tau$ values are
qualitatively very similar.  This is illustrated, in part, by
Fig.~\ref{f:cyltg:growth-time} where we show the the optimal growth envelope
(denoted by circles) at $\Rey=40$ and also the energy evolution from optimal
initial conditions corresponding to three quite different values of $\tau$.
While there are quantitative differences between the transient-response
curves, qualitatively they are similar, suggesting similar flow features and
dynamics are excited by initial conditions optimized across a large range of
\Tau\ values.  This holds for other values of \Rey, with the peak in the
response curves shifting to smaller times for smaller values of \Rey.  In the
spatiotemporal results which follow, we have opted to fix the \Tau\ at which
the optimal perturbations are computed, rather than having it vary with \Rey.
All optimal perturbations are for \Tau=20, as this provides a good choice over
the whole of the subcritical regime.

\begin{figure*}
\centering
\includegraphics[width=4.5in]{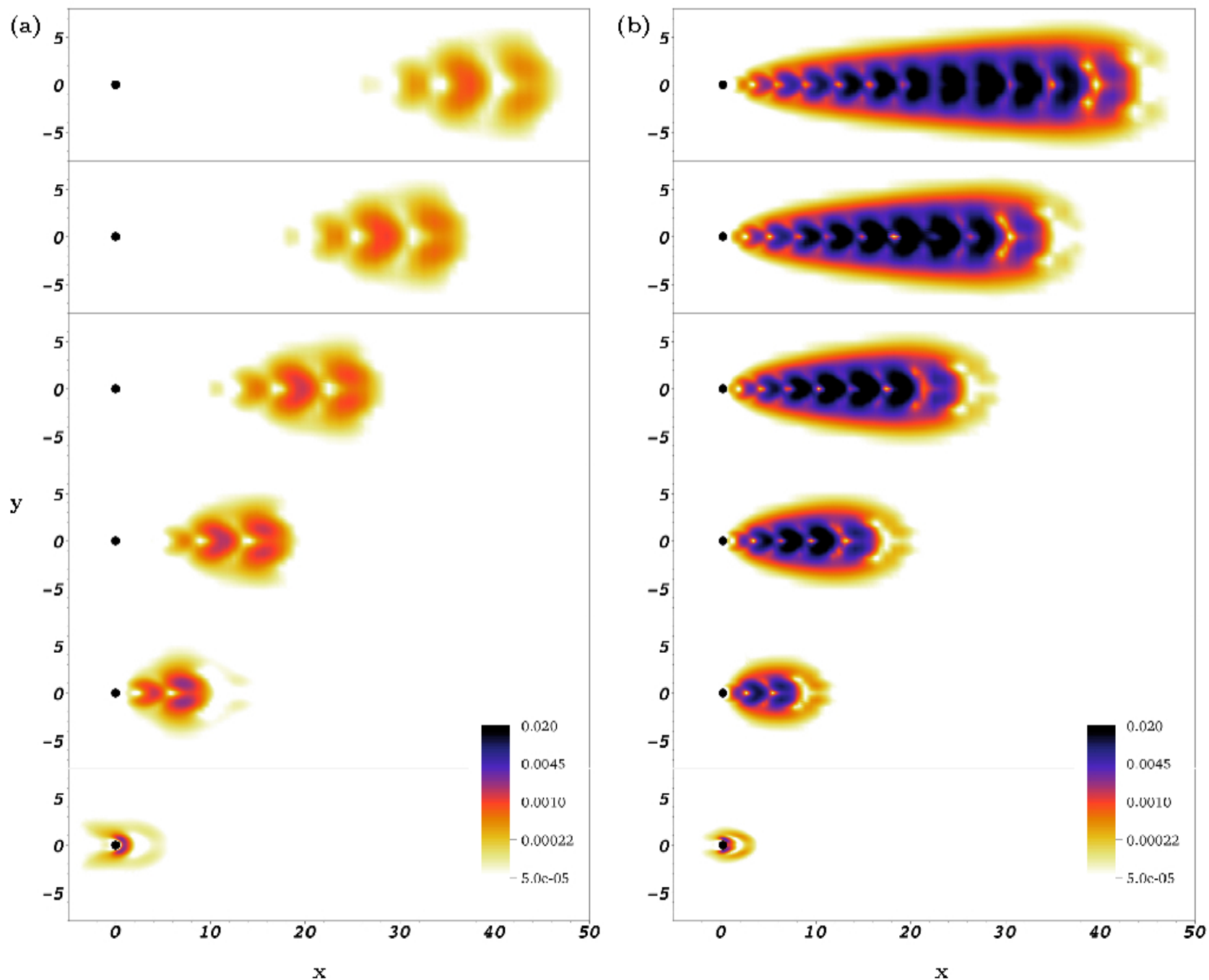}
\caption{(color online). Contours of energy showing the linear evolution of
  perturbations at \Rey=20 (left) and \Rey=40 (right).  The panels are
  snapshots at 10 time-unit intervals from $t=0$ (bottom) to $t=50$ (top).}
\label{f:visual-evolve}
\end{figure*}

\begin{figure*}
\centering
\includegraphics[width=4.5in]{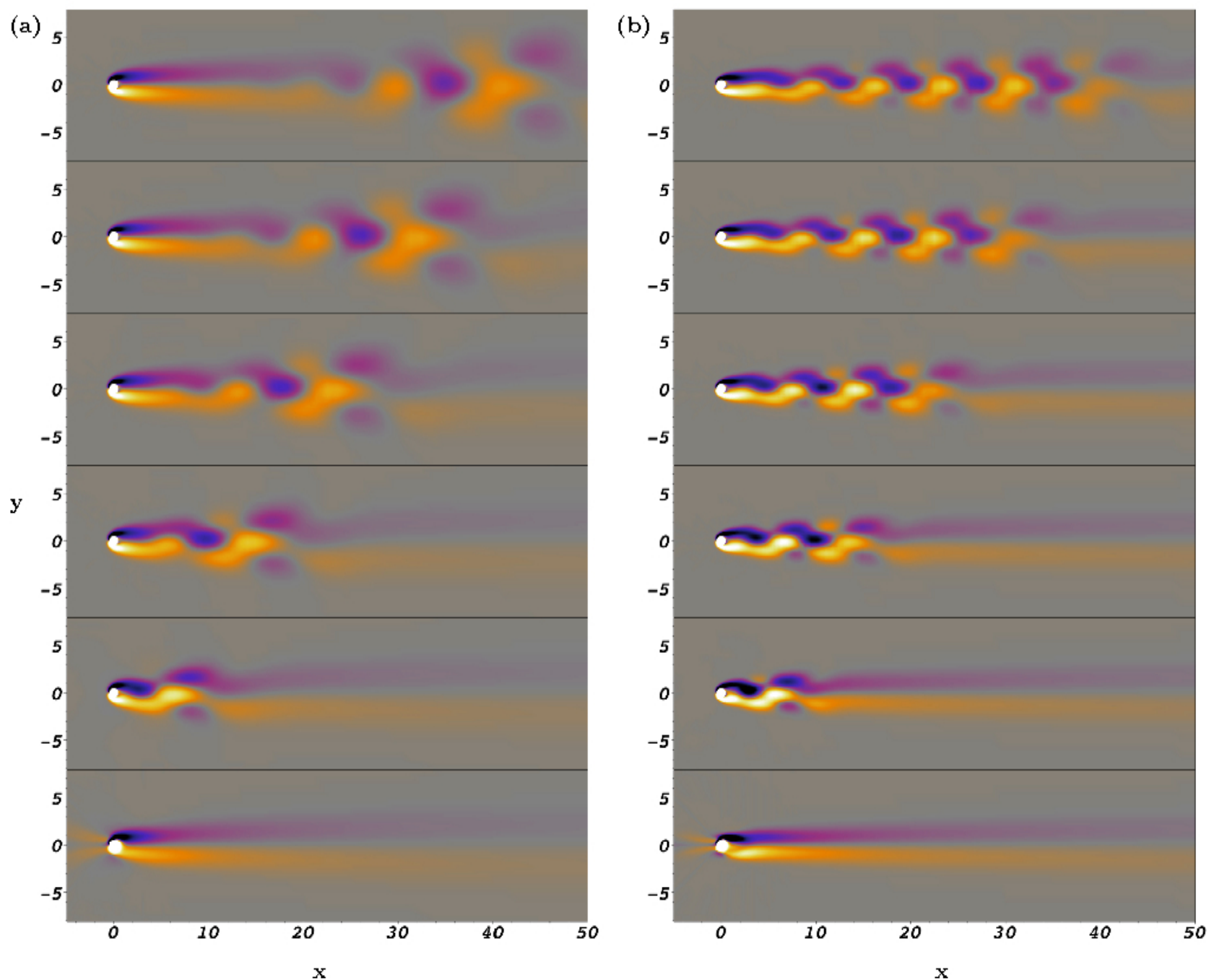}
\caption{(color online). Same evolution as in Fig.~\ref{f:visual-evolve},
  \Rey=20 (left) and \Rey=40 (right), but viewed in terms of vorticity.  The
  vorticity of a linear superposition of the base flow and the perturbation is
  shown at snapshots separated by 10 time units from $t=0$ (bottom) up to
  $t=50$ (top). The maximum vorticity is 1.5. }
\label{f:visual-evolve-vort}
\end{figure*}

Visualizations of the linear time evolution of perturbations $\up(t)$, at
\Rey=20 and \Rey=40 are shown in Figs.~\ref{f:visual-evolve} and
\ref{f:visual-evolve-vort}.  Perturbation energy, $|\up|^2/2$ is plotted in
Fig.~\ref{f:visual-evolve} with a fixed energy scale throughout the figure.
Figure~\ref{f:visual-evolve-vort} shows the same evolution, except in terms of
vorticity. Here we visualize not the vorticity in the perturbation field
itself, but in a superposition of the base flow and the perturbation, i.e. $\u =
\U + \epsilon \up$, where $\epsilon$ is chosen so that the resulting
superposition best resembles what one might find in an actual flow,
as for example, might be observed in experiments. In all cases only a portion
of the full computational domain is shown.

The bottom-most plots correspond to the optimal initial conditions $\up(t=0) =
\vv_1$.  One can see in the energy plot that the initial condition is more
localized to the cylinder at higher \Rey. In fact the initial condition
becomes quite broad spatially at low \Rey. In the vorticity plot one can see
the asymmetry of the combined flow introduced by the perturbation. The base
flow $\U$ is symmetric about the centerline, while the perturbation $\up(0)$
is antisymmetric.  Note, for the \Rey\ and \Tau\ values considered here, there
is no weak upstream tail in the initial conditions of the type shown in
Fig.~\ref{f:cyltg:opt-ic-inflow}(b), although weak upstream tails are found at
\Rey=40 for larger values of \Tau. These tails play no qualitative role in the
spatiotemporal dynamics.

The perturbation fields are evolved via the linearized Navier-Stokes
equations, Eq.~\eqref{e:cyltg:lnse}, and visualized every 10 time units.
The first obvious point is that in both cases the perturbation fields, or more
accurately the superposition of the perturbation fields and base flow,
resemble transitory \BvK\ vortex streets.
At $\Rey=20$, the initial perturbation develops into a packet of roughly two
wavelengths in streamwise extent and advects steadily downstream at a speed
slightly less than 1. The peak energy is reached at $t \simeq 27$ and
thereafter the energy decays quite gradually.  At \Rey=40, the leading edge of
the packet, and the streamwise location of the maximum of the response,
advects downstream at approximately the same speed as at \Rey=20. In this case
however, the evolving perturbation develops a long trailing series of sinuous
oscillations as the excited near-wake region undergoes slowly decaying
oscillations.  The streamwise wavelength of oscillations is smaller at \Rey=40
than at \Rey=20.  The peak energy at \Rey=40 is not reached until $t \simeq
60$, after the last plot shown. It is evident that the growth in the
integrated energy of the perturbation field is due both to an increase in the
maximum pointwise energy and also to a significant increase in the spatial
extent of the perturbation field. This second factor becomes increasingly
important as \Rey\ approaches $\Rey_c$.

\begin{figure*}
\centering
\includegraphics[width=4in]{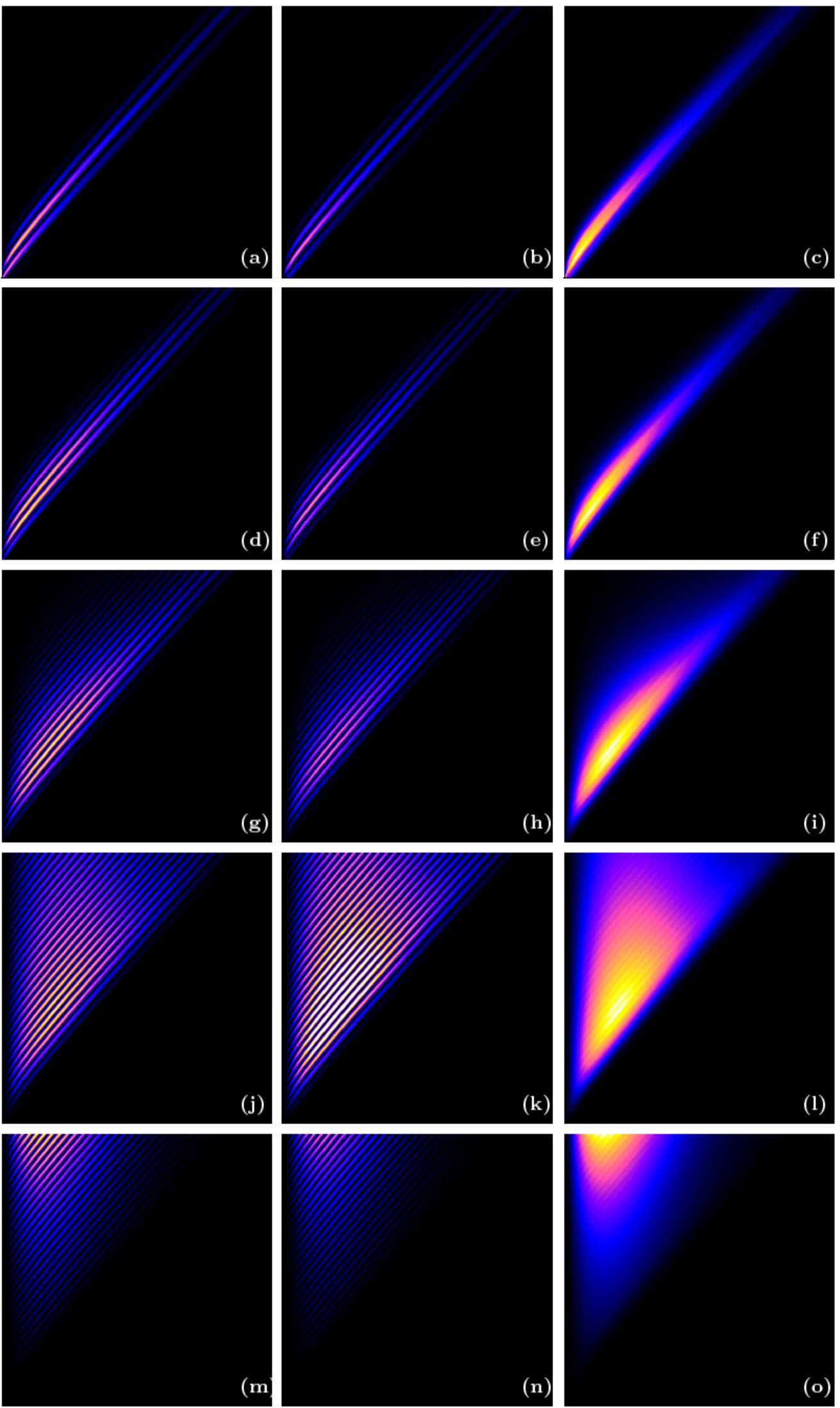}
\caption{(color online). Plots showing the space-time evolution of energy in
  the perturbation $\up$. Each tile covers $x$ between 0 and 125 (horizontal)
  and $t$ between 0 and 125 (vertical), and shows the energy of perturbations
  sampled on the centerline $y=0$.
  Each row corresponds to a different Reynolds number, specifically: $\Rey=20$
  (a,b,c), $\Rey=30$ (d,e,f), $\Rey=40$ (g,h,i), $\Rey=45$ (j,k,l), and
  $\Rey=50$ (m,n,o).
  For each Reynolds number we show the evolution of the dominant mode (left
  column), the first sub-dominant mode (center column), and a combination
  (right column) revealing the envelope of the perturbation as explained in the
  text.
  The scale of each row of tiles is normalized by the maximum energy in the
  right column over the space-time domain, with white corresponding to the
  highest energy and black to zero energy. }
\label{f:space-time}
\end{figure*}

To further highlight the spatiotemporal character of the evolving
perturbations, and their dependence on \Rey, we show in
Fig.~\ref{f:space-time} space-time diagrams covering a large range of both
space and time.  Within each tile, space is horizontal from $x=0$ to $x=125$
and time is vertical from $t=0$ to $t=125$. Hence unit speed, that of the
free-stream velocity, corresponds to $45^\circ$ in these plots.  Each row in
this figure corresponds to a particular \Rey, from \Rey=20 to \Rey=50.  The
left column shows the evolution of energy in the perturbation sampled on the
flow centerline, that is, contours of $\up(x,y=0,t)$, where $\up(t=0) = \vv_1$
is the optimal initial condition. For example, Fig.~\ref{f:space-time}(a)
shows the same perturbation as in Fig.~\ref{f:visual-evolve}(a).  The contour
scale varies from row to row and is set so that the maximum energy corresponds
to white and zero energy corresponds to black.

The center and right columns of Fig.~\ref{f:space-time} are explained as
follows. The center column is the evolution of the first sub-dominant optimal
mode, that is the evolution of $\up(t)$, where $\up(t=0) = \vv_2$.  The
sub-dominant mode and subsequent evolution are very similar to that of the
dominant mode. However, careful inspection shows that the sub-dominant mode is
spatially
phase shifted by a quarter wavelength with respect to the dominant mode. This
is seen as a half wavelength shift in Fig.~\ref{f:space-time} since a pair of
vortices (one wavelength) generates two peaks in the centerline energy.  The
pairing of perturbations has been observed and discussed
elsewhere~\cite{BlackburnBarkleySherwin:2008, CantwellBarkleyBlackburn:2010}
as a manifestation of streamwise symmetry breaking such that modes come in
near pairs with similar, but not identical, dynamics. 
The importance of this second,
phase-shifted mode is that from the pair of modes we can easily construct an
approximate energy envelope eliminating the fast oscillations associated with
vortex shedding. This is shown in the third column where we plot $E=E_1+\alpha
E_2$, where $E_1$ and $E_2$ are the energy of the dominant (left column) and
sub-dominant (middle column) perturbation fields. We choose $\alpha$ so that
the peak energy of the sub-dominant mode matches that of the dominant mode. As
one can see this nearly eliminates the fast oscillations throughout the
space-time plot of $E$.


The dynamics seen at \Rey=20, \Rey=30, and \Rey=40 are quite similar. There is
an increase in energy (both peak energy and spatial extent) followed by a
decrease with the long-term dynamics being a weak wave packet propagating and
decaying downstream.  The effects of varying \Rey\ in the regime are those
already noted: there is a decrease in wavelength and an increase near-wake
oscillations with increasing \Rey. 

The behavior at \Rey=45 is, however, qualitatively different from that seen at
\Rey=40 and below, even though \Rey=45 is still in the subcritical regime. The
perturbation at long times does not have a maximum at some downstream location
set by how long the perturbation has evolved. Instead the maximum is located at
a finite streamwise location. This is due to the fact that at long times the
perturbation must evolve to the least stable wake eigenmode and there is a
qualitative change in the spatial structure of this eigenmode at
$\Rey\simeq42$ (also noted by Giannetti and Luchini who give $\Rey\simeq43$).
Below $\Rey=42$ the leading eigenmode is exponentially growing downstream and
hence appears localized to the downstream computational boundary. Above
$\Rey=42$ the leading eigenmode has a maximum at finite streamwise position
with exponential decay far downstream. The location of the maximum decreases
as a function of \Rey\ and is at about $x=34.6$ for \Rey=45.  This phenomenon
is well-known and understood in other systems, \eg~\cite{
Barkley:1992, SandstedeScheel:2000, WheelerBarkley:2006}. 
In these systems, the switch from downstream growth to downstream decay of an
eigenfunction occurs when the corresponding eigenvalue crosses the essential
spectrum. The essential spectrum, in turn, is the continuous eigenvalue
spectrum associated with the far-field part of the system. It might be of some
interest in the future to investigate these issues for the cylinder wake. 

For completeness we also show the evolution at $\Rey=50$, slightly into
unstable regime. Although somewhat masked by the fact that the perturbation is
growing, the spatial structure of the mode at long time is not very different
from that at $\Rey=45$. The perturbation has a maximum at about $x=19.4$
followed by exponential downstream decay, matching that of the leading
eigenmode from the stability analysis at \Rey=50.

\subsection{3D Energy Growth}

\begin{figure}
\centering
\includegraphics{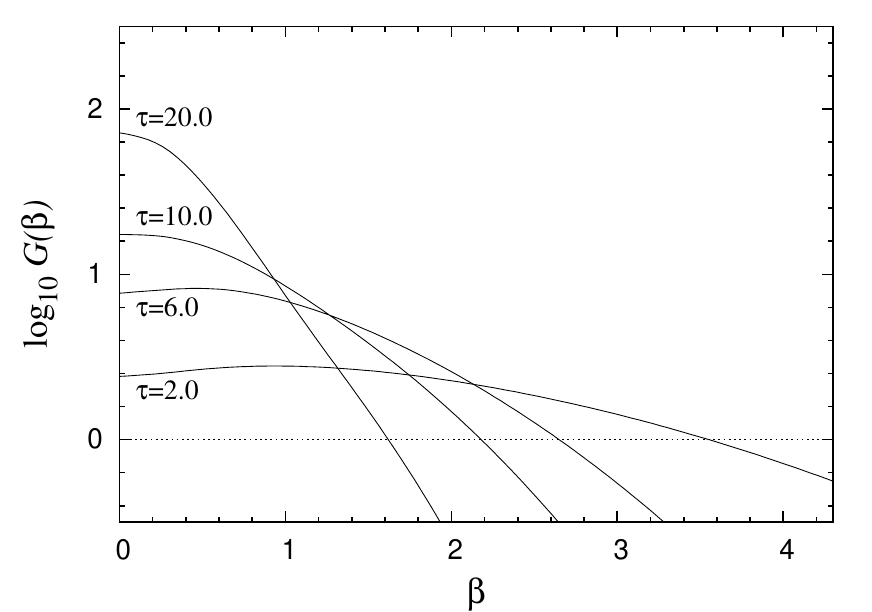}
\caption{Optimal growth as function of spanwise wavenumber $\beta$ at
  \Rey=40. $\beta=0$ is dominant for long time horizons, but higher $\beta$
  may provide slightly larger growth at short time horizons.}
\label{f:growth-wave}
\end{figure}

\begin{figure}
\centering
\includegraphics{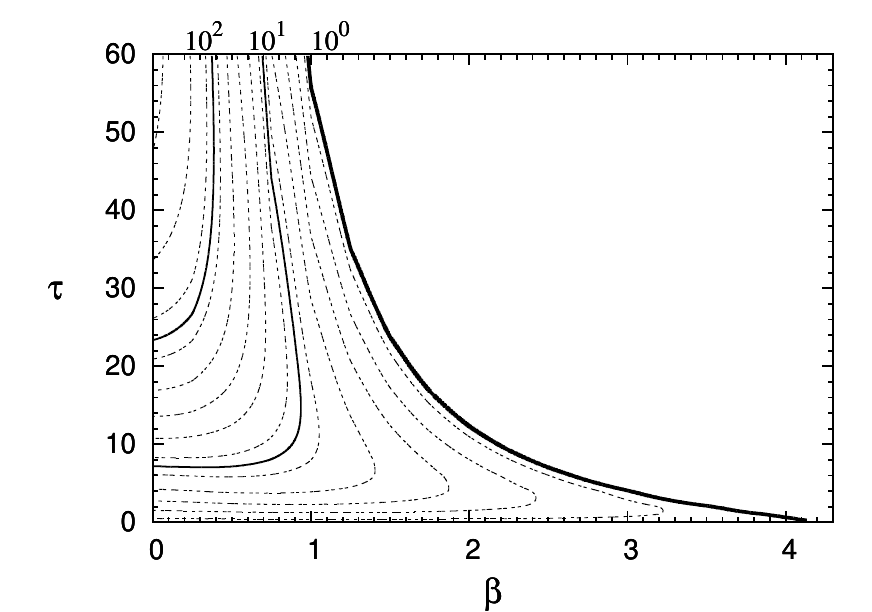}
\caption{Contour plot of optimal growth at \Rey=40. The thicker black line
  denotes the contour of no growth.}
\label{f:growth-wave-contour}
\end{figure}

We consider here briefly the energy growth of 3D perturbations, mainly to show
that 3D effects are unimportant.  The spanwise wavenumber $\beta$ of
perturbations, Eq.~\eqref{eq:3Dpert}, becomes an additional parameter to
vary. We shall fix the Reynolds number at $\Rey=40$.  Optimal growth curves
over a range of spanwise wavenumbers, at representative values of $\tau$, are
plotted in Fig.~\ref{f:growth-wave} and growth contours in the $\beta$-$\tau$
plane are shown in Fig.~\ref{f:growth-wave-contour}.  The thicker line in
Fig.~\ref{f:growth-wave-contour} denotes the no-growth contour and energy
growth occurs only to the left of this contour.  Except for small values of
$\tau$, the growth of 2D perturbations ($\beta=0$) greatly dominates the
growth of 3D perturbations.

For short time horizons, (we estimate $\tau \lesssim 8.0$), the largest
possible growth is found at nonzero wavenumbers and the range of active
wavenumbers increases considerably as $\tau$ approaches zero.  This shift to
high-wavenumber modes at short time horizons occurs in other shear
flows~\cite{SchmidHenningson:1994, BlackburnSherwinBarkley:2008,
  CantwellBarkleyBlackburn:2010}, but we are unaware of any explanation for
this phenomenon.  This does not seem important in any practical sense because
the overall response of such modes is very small indeed.  We have not
investigated other values of $\Rey$ in detail, but the unsurprising result is
that 2D modes dominate the transient response in the subcritical wake.

\section{Summary and Discussion}

We have studied the subcritical response of the cylinder wake by accurately
computing the optimal energy growth throughout the subcritical regime. We have
treated at some length the numerical domain requirements for accurate
computations within the subcritical region. The results themselves show that
energy growth first occurs as low as $\Rey \simeq 2.2$, below the onset of
separation at $\Rey \simeq 6.2$.  Over most of the subcritical regime the
maximum energy amplification increases approximately exponentially in the
square of $\Rey$.  This super-exponential dependence on $\Rey$ is even faster
than the exponential dependence commonly observed in other separated flows
~\cite{BlackburnBarkleySherwin:2008, BlackburnSherwinBarkley:2008,
  CantwellBarkleyBlackburn:2010}. However, the maximum growth in the cylinder
wake never reaches the extremely large values since the wake becomes linearly
unstable at a relatively low \Rey\ where the maximum energy growth is about
6800.

We have considered the structure of the optimal transient dynamics. The
evolving perturbations are of the form of transitory \BvK\ vortex streets. At
lower \Rey\ wave packets of only a few wavelengths are formed which propagate
downstream. As \Rey\ increases the packets extend in length due to the slow
decay of oscillations in the near wake. At $\Rey\simeq42$ the spatial
structure of the response at long times switches from exponentially increasing
downstream to exponentially decaying downstream so that at about $\Rey=42$ the
response at long times has a maximum at a finite streamwise location.  Finally,
at $\Rey = \Rey_c \simeq 46.6$ the wake becomes linearly unstable.

It is of interest to relate our results to the understanding of subcritical
dynamics arrived at by local stability analysis, \eg
\cite{Monkewitz:1988,YangZebib:1989, HannemannOertel:1989, HuerreMonkewitz1990,
  Delbende1998, Pier:2002, Chomaz:2005} and references therein.  In brief,
from sectional stability analysis of wake profiles, the picture of the
subcritical region is as follows. Below $\Rey \sim 5$ the wake is everywhere
stable. Above $\Rey \sim 5$ there is a region of convective instability behind
the cylinder and at $\Rey \sim 25$ a pocket of absolute instability appears
within the region of convective instability. The size of the absolute pocket
grows with \Rey\ and is thought to be responsible for the actual instability
occurring at $\Rey_c$, although prediction of the transition point has eluded
local analysis.

In reality, there are two qualitative changes within the subcritical regime:
the onset of transient growth at \Rey=2.2 and the switch from downstream
growth to downstream decay of transient structures at $\Rey\simeq 42$,
associated with a corresponding change in the structure of eigenmodes.  It seems
that the first of these, the onset of transient energy growth, could be
connected with the first appearance of a local convective instability. A local
pocket of convective instability would indeed correspond to transient response
in a global setting. Moreover, the \Rey\ values for the two event are
reasonably close.  
This would appear to corroborate the picture first proposed by Cossu and
Chomaz \cite{CossuChomaz:1997} in the context of the Ginzburg-Landau equation.
In this picture, 
one can understand the transient energy
growth as arising from perturbations traveling through a local region of
instability, where they are amplified, followed by advection into the stable
downstream wake, where they decay. 
We caution, however, that the cylinder wake
is highly non-parallel in the near wake region and it would probably be a
mistake to connect the transient response and the parallel-flow
analysis in too much detail. 

There is nothing in the actual transient response corresponding to
the local opening of the absolute pocket at $\Rey \sim 25$, but neither is
there expected to be~\cite{Delbende1998}. We have clearly shown an uneventful
evolution of the transient response between $\Rey = 20$ and $\Rey = 30$, and
in fact up to $\Rey = 40$.  We are unaware of any local analysis of the
cylinder wake that predicts the shift from growth to decay of modes at
$\Rey\simeq42$, and this might be interesting to investigate in the future.

There is another way to view the relationship between our study and concepts
of convective and absolute instability.  This is also closely related to some
past and ongoing experimental studies~\cite{LeGalCroquette:2000,
  Maraisetal:2010}.  While convective and absolute instability are strictly
defined for streamwise homogeneous flows, which the cylinder wake is not, the
change in the linear response at $\Rey_c$ has the essential character of the
transition from convective to absolute instability and it commonly referred to
using these terms.  One sees this in our Fig.~\ref{f:space-time} where the
subcritical cases, Figs.~\ref{f:space-time}(c), \ref{f:space-time}(f),
\ref{f:space-time}(i) and \ref{f:space-time}(l) have the character of
convective instability: initial perturbations lead to wave packets that advect
downstream such that even though a perturbation grows (for some time) it is
simultaneously advected quickly downstream.  The supercritical case
\ref{f:space-time}(o) has the character of absolute instability where
perturbations grow at fixed streamwise locations.  LeGal and Croquette present
nice streakline visualizations of the transient wake, qualitatively similar to
what is shown in our Fig.~\ref{f:visual-evolve-vort}, and discuss this as
evidence of convective instability in the cylinder wake prior to the onset of
sustained oscillations.  Marias {\em et al.} use particle image velocimetry
(PIV) to obtain more quantitative measures of the subcritical response
generated by rotary motion of the cylinder. In particular they measure front
velocities and study how these behave as $\Rey_c$ is approached.  Marias {\em
  et al.} also extract integrated energy from PIV data. Transient
amplification is indeed observed, followed by exponential decay.  However, due
to the fact that experimental perturbations are introduced by cylinder
rotation, and not from the optimal initial conditions studied here, quantitative
comparisons are not presently possible, but may be pursued in the future.

Finally, we conclude with the issue of numerical accuracy.  Our study has
highlighted the importance of ensuring the numerical convergence of the
computational domain. Transient growth problems in open flows with
inflow-outflow boundary conditions are particularly susceptible to
deficiencies in the extent of the computational domain. This is true not only
in the downstream region, but in the cross-stream and especially the inflow
dimensions. It is well known that for external flows enforcing boundary
conditions too close to a body can lead to deformation of the underlying basic
flow~\cite{Fornberg:1980,LecointePiquet:1984,StrykowskiHannemann:1991,
AnagnostopoulosIliadisRichardson:1998}.
Accurate resolution of perturbation fields for transient growth problems can 
impose yet more severe requirements.  The cylinder wake is a prime example
of a flow in which the requisite domain can be far greater for transient
growth computations than for other types of calculations.

\begin{acknowledgments}

Computing facilities were provided by the UK Centre for Scientific Computing
of the University of Warwick.  DB gratefully acknowledges support from the
Leverhulme Trust and the Royal Society.

\end{acknowledgments}

\appendix*
\section{}
\label{sec:appendix}


In this appendix we present convergence results for base-flow calculations,
stability calculations, and polynomial order.

\begin{figure}
\centering
\includegraphics{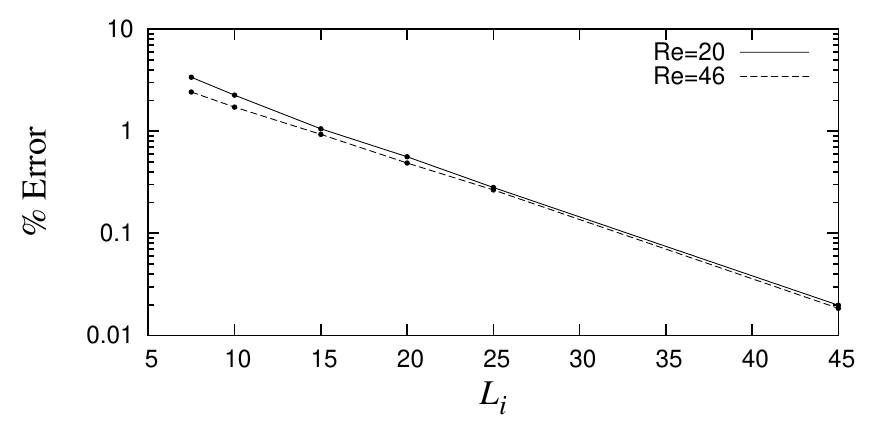}\\
\includegraphics{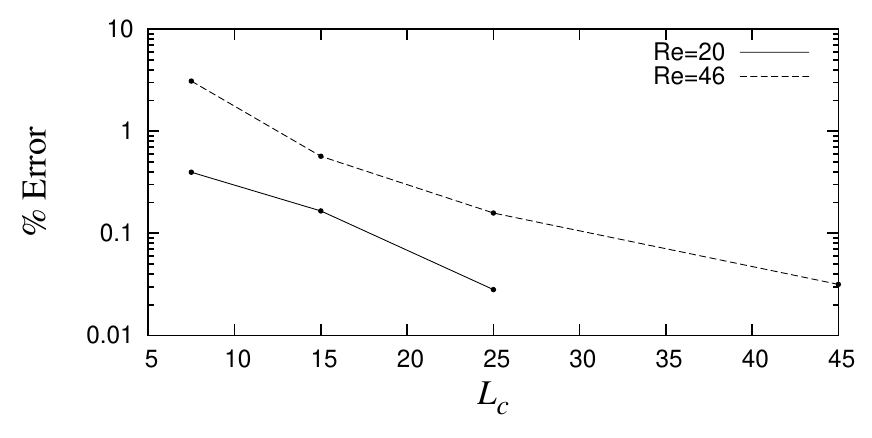}
\caption{Convergence of the base flow stagnation point with mesh
  dimensions. Points indicate the computed values. 
In (a) $L_c$ is fixed at 25 and in (b) $L_i$ is fixed at 25. }
\label{f:cyltg:base-stag}
\end{figure}

Base flow convergence is assessed through two indicators: the position, $x_s$,
of the stagnation point marking the end of the recirculation region and
velocity profiles just downstream of the cylinder.
Figure~\ref{f:cyltg:base-stag} summarizes the convergence of the stagnation
point with mesh dimension.  The stagnation point is not present at $\Rey=5$,
and consequently this case does not appear.  Percentage errors are relative to
the calculation using $L_i=65$ and $L_c=65$, respectively.  The stagnation
point is seen to be highly converged, as a function of domain dimensions, for
$L_i=L_c=45$.

We may also compare the values of $x_s$ with values reported in previous
studies~\cite{DennisChang:1970, CoutanceauBouard:1979, Fornberg:1980,
  ZielinskaGoujonDurandDusekWesfreid:1997, YeMittalUdaykumarShyy:1999, 
  Pier:2002,GiannettiLuchini:2007}.
Consistent with other studies, we find for $\Rey \ge 6.2$ the stagnation point
obeys $$ x_s \simeq 0.5 + 0.067 (\Rey - 6.2),
$$ 
with specific converged values: $x_s = 1.422$ at $\Rey=20$ and $x_s = 2.762$
at $\Rey=40$.  These agree very will with recent computational studies by
Giannetti and Luchini~\cite{GiannettiLuchini:2007} and
Ye~\etal~\cite{YeMittalUdaykumarShyy:1999}.


\begin{figure}
\centering
\includegraphics{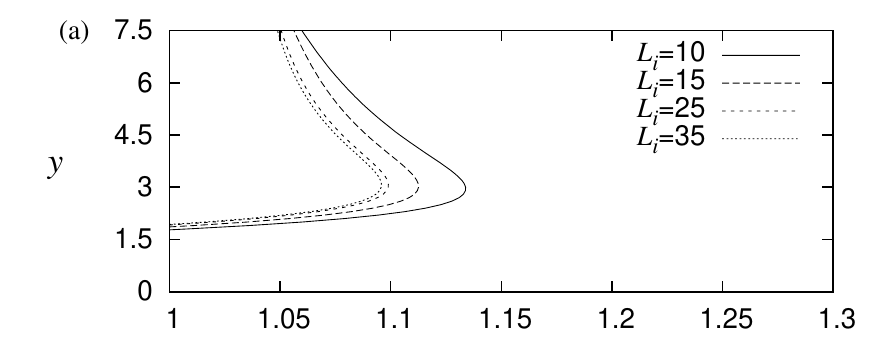}\\
\includegraphics{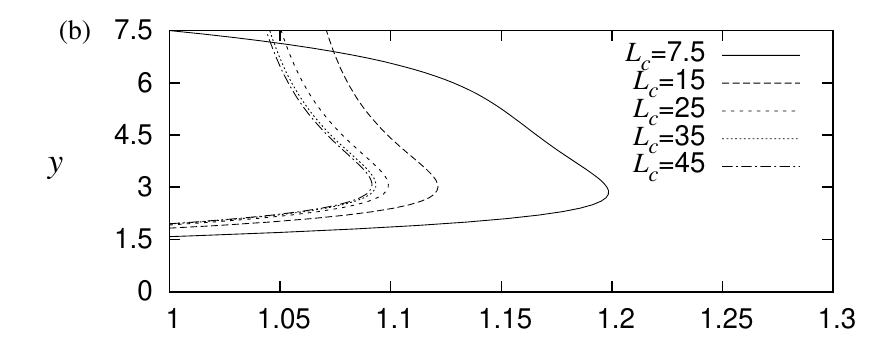}\\
\includegraphics{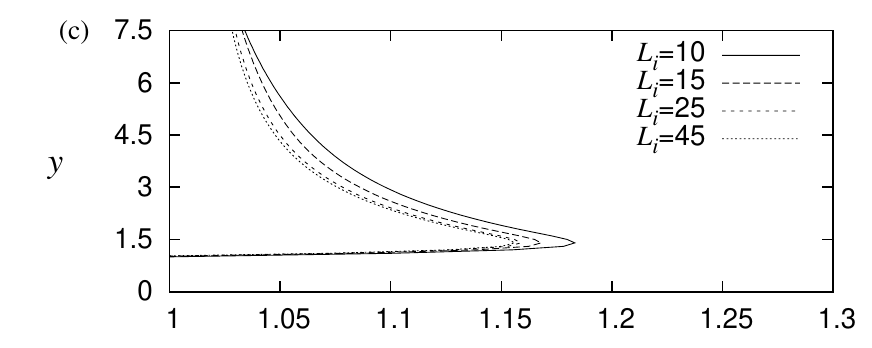}\\
\includegraphics{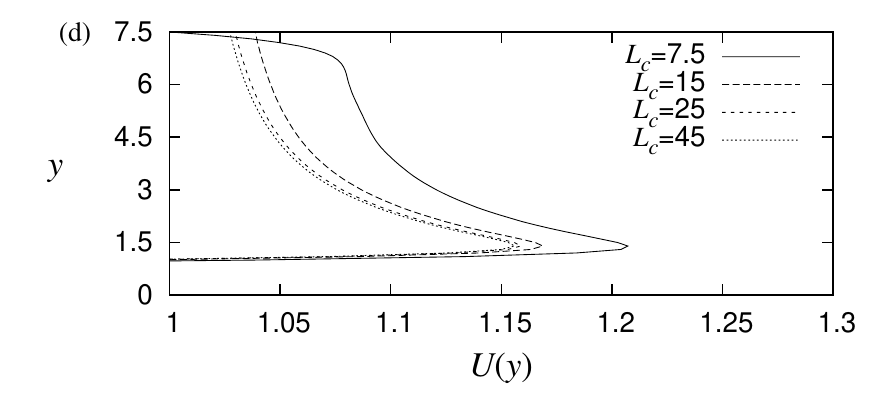}
\caption
{Streamwise velocity profiles of base flows at \Rey=5, $(a)$ and $(b)$, and at
  \Rey=46, $(c)$ and $(d)$, showing variation with $L_i$ and $L_c$. For
  variations in $L_i$, we fix $L_c=25$. For variations in $L_c$, we fix
  $L_i=25$.}
\label{f:cyltg:base-profile}
\end{figure}

Examination of streamwise velocity profiles is found to provide a more
detailed view of base-flow distortion due to the finite-size
effects. Figure~\ref{f:cyltg:base-profile} shows velocity profiles at location
$x=3$. Only a limited cross-stream range in $y$ is shown in the vicinity of
where the streamwise velocity reaches its maximum, as this is where the
effects of domain confinement are most pronounced.  Constriction of the
cross-stream mesh leads to an especially inaccurate calculation of the base
flow, particularly at low \Rey, while the effect of restricted inflow length
is less significant in general [Fig.~\ref{f:cyltg:base-profile}(a) and (c)].
In any case, the base flow is again seen to be highly converged, as a function
of domain dimensions, for $L_i=L_c=45$.

\begin{figure}
\centering
\includegraphics{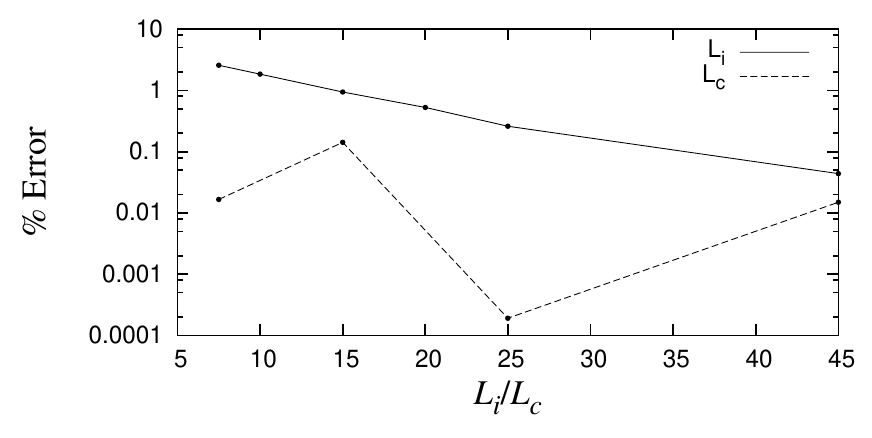}
\caption
{Convergence of critical Reynolds number $\Rey_c$ with mesh size.
Points indicate the computed values. }
\label{f:cyltg:linear}
\end{figure}

The dependence of the linear stability calculations on domain size is examined
through determination of the critical Reynolds number, $\Rey_c$, on different
domains. For each domain, we compute the base flow and the eigenvalues at
$\Rey = 43, 44, 45,$ and $46$. From these we extrapolate to find $\Rey_c$
where the real part of the leading eigenvalue crosses zero. 
The results are shown in Fig.~\ref{f:cyltg:linear}, where as before we report
percentage error in the value of $\Rey_c$ with respect to the value obtained
using $L_i=65$ and $L_c=65$, respectively.  Interestingly, one sees very little
effect of cross-stream restriction here. In any case, $\Rey_c$ is well
determined for $L_i = L_c =45$, with an error of less than $0.1\%$.

We may also compare directly the value we obtain for $\Rey_c$ with that
obtained in other studies.  To three significant figures, with $L_i = L_c
=45$, we find
$$
\Rey_c = 46.6.
$$
This value agrees to within half a percent with recent stability calculations
by Giannetti and Luchini~\cite{GiannettiLuchini:2007} and Marquet
\etal~\cite{MarquetSippLaurent:2008} who quote values of $\Rey_c=46.7$ and
$\Rey_c=46.8$, respectively.

Finally, having set the overall mesh dimensions, we consider the convergence
of computations with respect to the polynomial order of the spectral-element
expansion.  The polynomial order is chosen to ensure there is the necessary 
refinement to resolve the finest characteristics of the flow at the highest
Reynolds number under consideration.  Base flow and subsequent transient growth
calculations at $\Rey=46$ have been carried out for a range of polynomial orders
as summarized in Table~\ref{t:cyltg:converge-poly}. A polynomial order of $P=6$
is found to be sufficient and is used for all results reported in
Sec.~\ref{sec:cyltg:tg}.

\begin{table}
\begin{ruledtabular}
\begin{tabular}{cc}
Order & 	$G(\tau=20)$ \\
\hline \\
3 &			138.91  \\
4 &			108.29  \\
5 &			156.29  \\
6 &			156.20  \\
7 &			156.19  \\
8 &			156.19  \\
\end{tabular}
\end{ruledtabular}
\caption
{Convergence of optimal growth results with polynomial order on the mesh
  $L_i=45$, $L_c=45$, $L_o=125$. The base flow and optimal growth $G(\tau=20)$
  at $\Rey=46$ are both computed for the polynomial orders indicated.}
\label{t:cyltg:converge-poly}
\end{table}

\end{document}